\documentclass[aps,preprint,eqsecnum,groupedaddress,preprintnumbers]{revtex4}

\usepackage{graphicx}
\usepackage{bm}
\usepackage{amsmath}
\usepackage{amssymb}
\usepackage{amsfonts}
\usepackage[dvips]{color}
\newcommand{\half}{\frac{1}{2}}

\newcommand{\der}{\partial}

\newcommand{\Tr}{\mbox{\rm Tr}}
\newcommand{\dsl}{\partial\kern-0.55em\raise 0.14ex\hbox{/}}

\begin{document}

\preprint{KYUSHU-HET-117}

\title{Apparently non-invariant terms of nonlinear sigma models in
lattice perturbation theory}

\author{Koji Harada}
\email{harada@phys.kyushu-u.ac.jp}
\affiliation{Department of Physics, Kyushu University\\
Fukuoka 810-8560 Japan}
\author{Nozomu Hattori}
\email{hattori@higgs.phys.kyushu-u.ac.jp}
\affiliation{Department of Physics, Kyushu University\\
Fukuoka 810-8560 Japan}
\author{Hirofumi Kubo}
\email{kubo@higgs.phys.kyushu-u.ac.jp}
\affiliation{Department of Physics, Kyushu University\\
Fukuoka 810-8560 Japan}
\author{Yuki Yamamoto}
\email{yamamoto@phys-h.keio.ac.jp}
\affiliation{Hiyoshi Department of Physics, Keio University \\
Yokohama 223-8521 Japan}

\date{\today}

\begin{abstract}
 Apparently non-invariant terms (ANTs) which appear in loop diagrams for
 nonlinear sigma models (NLSs) are revisited in lattice perturbation
 theory. The calculations have been done mostly with dimensional
 regularization so far. In order to establish that the existence of ANTs
 is independent of the regularization scheme, and of the potential
 ambiguities in the definition of the Jacobian of the change of
 integration variables from group elements to ``pion'' fields, we employ
 lattice regularization, in which everything (including the Jacobian) is
 well-defined. We show explicitly that lattice perturbation theory
 produces ANTs in the four-point functions of the ``pion'' fields at
 one-loop and the Jacobian does not play an important role in generating
 ANTs.
\end{abstract}


\maketitle

\section{Introduction}

There has been a long time since apparently chiral non-invariant,
divergent contributions were noticed in the loop calculations of
nonlinear sigma (NLS) models. It is important to realize that there are
two kinds of such contributions. The first kind, which produces the mass
term of the pion field, leads to the violation of the soft pion
theorem. The second kind is more subtle. It does not violate the soft
pion theorem and is claimed to vanish on-shell. It is well understood
that the first kind contributions are canceled by those from the
Jacobian~\cite{Charap:1971bn, Honerkamp:1971va, Gerstein:1971fm}. (They
had been overlooked at that time.) In the dimensional regularization,
this kind of non-invariant contributions are absent; it is consistent
with the absence of the nontrivial Jacobian in this regularization
scheme. As for the second kind contributions, although they have been
discussed in the literature~\cite{Tataru:1975ys, Honerkamp:1971sh,
Kazakov:1976tj, deWit:1979gw, Appelquist:1980ae}, there still seems to
be unclear points, on which we are going to discuss in this paper.

The prescriptions of how to avoid the second kind have been
proposed. T\u{a}taru~\cite{Tataru:1975ys} showed, using dimensional
regularization, that the second kind contributions are proportional to
the (classical) equations of motion and do not contribute to the
S-matrix, following the argument by
't~Hooft~\cite{'tHooft:1973us}. Honerkamp~\cite{Honerkamp:1971sh} and
Kazakov, Pervushin, and Pushkin~\cite{Kazakov:1976tj} proposed to use
the background field method. This is essentially to modify the
theory. Appelquist and Bernard~\cite{Appelquist:1980ae} pointed out that
a field redefinition removes such contributions. The most popular and
practical method is to consider not the pion field but the
currents~\cite{Gasser:1983yg, Gasser:1984gg}.  In recent papers Ferrari
\textit{et al.}~\cite{Ferrari:2005ii, Ferrari:2005va, Ferrari:2005fc,
Bettinelli:2007zn} reconsidered the renormalization problem emphasizing
the symmetry point of view, heavily relying on the Ward-Takahashi
identities, and gave the subtraction procedure consistent with
them. They claim that the use of the dimensional regularization, in
which the tadpole contributions are absent, is essential.



In this paper, we instead use lattice regularization for the following
reasons: (i) Since everything is well defined in the lattice
regularization, it is obvious that there is no source of the violation
of chiral symmetry (up to a ``spurion'' mass term), if we start with a
symmetric partition function. This fact is important for establishing
that chiral symmetry is not lost despite the appearance of ANTs. Hence
the name; they \textit{do not} violate chiral symmetry, though they
\textit{appear} to be non-invariant. (ii) In the case of the first kind
contributions, the Jacobian plays an essential role. It is interesting
to see if the Jacobian plays any role for the second kind. The logarithm
of the Jacobian is proportional to $\delta^4(0)$, thus in the
dimensional regularization it is trivially set to zero, while in other
continuous regularization schemes it is ill-defined. In the lattice
regularization, on the other hand, it is regularized and well-defined,
so that one can carefully examine the effects of the Jacobian. One might
suspect that the (naive) Jacobian is actually the latent source of the
violation of chiral symmetry, and that a properly defined Jacobian
should contain momentum-dependent terms in order for the theory to be
chiral invariant, which eventually cancel the ANTs produced by loop
diagrams.  It is therefore important to see what happens with the
well-defined, momentum-independent Jacobian in the manifestly chiral
invariant theory. (iii) Lattice regularization is completely different
from dimensional regularization.  It is therefore useful to see if the
existence of ANTs is independent of the regularization scheme. To our
best knowledge, ANTs in four dimensions have never been calculated by
using lattice regularization in the literature. (In 2+$\epsilon$
dimensions, Symanzik~\cite{Symanzik:1983gh} obtained ANTs in the lattice
regularization.)

The purpose of this paper is to establish the existence of ANTs in the
lattice perturbation theory at one-loop preserving chiral symmetry
manifestly. This implies that ANTs are compatible with chiral symmetry.
We also see that the Jacobian does not play an important role in
generating ANTs and that the appearance of ANTs is independent of
regularization schemes.

Our calculation is a straightforward generalization of Shushpanov and
Smilga~\cite{Shushpanov:1998ms}, who calculated only the self-energy
contributions.  We consider the four-point (amputated) Green functions
at low momenta ($p \ll 1/a$) to order $\mathcal{O}(p^4)$ at one-loop
level. A mass term is introduced in order to regularize the IR
singularities. Unlike the self-energy calculation, the IR regularization
with the mass term plays an important role for the calculations of the
four-point functions. We find that the divergent part of it contains
ANTs, which cannot be removed by a symmetric counterterms. We also find
that the Jacobian does not play an essential role. The ANTs vanish on
the mass shell.

In the next section, we establish the existence of the ANTs by an
explicit one-loop calculation. In Sec.~\ref{sec:conclusion}, we summarize
the results and give discussions. Appendix~\ref{sec:formulae} contains
some integration formulae.

\section{Lattice perturbation theory}
\label{sec:lattice}

\subsection{Setup}

In this section, we give an explicit one-loop calculation for the
four-point amputated Green function in the $SU(2)\times SU(2)$ NLS model
in four dimensions. In the NLS as an effective theory there are
infinitely many terms with increasing number of derivatives.  We are
however interested only in whether there arises an ANT of
$\mathcal{O}(p^2)$ or of $\mathcal{O}(p^4)$ at one-loop level. (Note
that, unlike the dimensional regularization, there are contributions of
$\mathcal{O}(p^2)$ from the one-loop diagrams in the lattice
regularization.) To see this, we will consider the one-loop
contributions only with vertices of $\mathcal{O}(p^2)$ and examine
whether the contributions of $\mathcal{O}(p^2)$ and of
$\mathcal{O}(p^4)$ can be absorbed in the symmetric terms. There may
be other ANTs involving higher derivative vertices, but they are not
related to the lower order contributions by the symmetry, and cannot
cancel the ANTs that may arise to this lowest order.

In the continuum, the action of $\mathcal{O}(p^2)$ is given by
\begin{equation}
 \mathcal{L}_2=\frac{F^2}{4}\Tr
  \left(
   \der_\mu U^\dagger \der_\mu U
  \right)
  -\frac{F^2m^2}{4}\Tr
  \left(
   U + U^\dagger
  \right),
  \label{lowest}
\end{equation}
where $U$ is an $SU(2)$-valued field and $F$ is the coupling
constant. (In the dimensional regularization, it is the pion decay
constant in the chiral limit.) We also introduce the mass term to
regularize the IR singularities.

On the hypercubic lattice with $a$ being the lattice constant, the
action may be written as
\begin{equation}
 S^{lat}_2[U]=\frac{F^2a^2}{4}\sum_n
  \bigg[
   \sum_{\mu} \Tr
   \left(
    2-U_{n}^\dagger U_{n+\mu}-U^\dagger_{n+\mu}U_n
   \right)
   -m^2a^2 \; \Tr
   \left(
    U_n^\dagger + U_n
   \right)
  \bigg],
\end{equation}
which is obtained by the simple replacement,
\begin{equation}
 \der_\mu U(x) \rightarrow (U_{n+\mu}-U_n)/a.
\end{equation}
There are many other discretization methods, but the choice does not
make a crucial difference in the following discussions so that we stick
to this simplest choice.

The partition function is given by
\begin{equation}
 Z=\int \prod_n DU_n \; e^{-S_2^{lat}[U]},
\end{equation}
where $DU_n$ stands for the invariant measure under the global
$SU(2)_L\times SU(2)_R$ transformations,
\begin{equation}
 U_n \rightarrow g_L U_n g_R^\dagger,
\end{equation}
where $g_L$ and $g_R$ are $SU(2)_{L,R}$ elements. Note that if the mass
term is treated as a ``spurion'' field~\cite{Gasser:1983yg}, and
transformed properly, the theory is manifestly invariant under
$SU(2)_L\times SU(2)_R$.

We introduce pion fields to do perturbation theory. We employ the
following parameterization,
\begin{equation}
 U_n=\sigma_n +i\pi^a_n \tau^a /F,\quad
  \sigma_n=\sqrt{1-\left(\pi^a_n\right)^2/F^2}.
\end{equation}
There are of course other parameterizations. But the main results are
independent of the choice.

In terms of the pion fields, the measure is written as
\begin{equation}
 \prod_n DU_n = e^{-S_{Jacob}^{latt}}\prod_{n,a} D\pi_n^a,
\end{equation}
with~\cite{Boulware:1970zc}
\begin{equation}
 S_{Jacob}^{lat}=-\frac{1}{2}a^4\sum_n\frac{1}{a^4}\Tr \ln 
  \left[
   \delta_{ab}+\frac{\pi_n^a \pi_n^b}{F^2-(\pi_n^c)^2}
  \right].
\end{equation}
Note that the $\delta^4(0)$ is regularized as $1/a^4$ on the lattice. It
is important to note that the vertices from the Jacobian is momentum
independent.

Expanding $S_2^{latt}$ and $S_{Jacob}^{latt}$ in terms of the pion fields
$\pi$, we obtain
\begin{eqnarray}
 S^{lat}_2&=&\frac{a^2}{2}\sum_n
  \left[
   \sum_\mu\left(\pi^a_{n+\mu}-\pi^a_n\right)^2
   +m^2a^2\left(\pi^a_n\right)^2
  \right]
  -\frac{a^2}{4F^2}\sum_{n,\mu}
  \left(\pi_n^a\right)^2\left(\pi^b_{n+\mu}\right)^2
  \nonumber \\
 &&+\frac{a^2}{8F^2}\left(m^2a^2+8\right)\sum_n
  \left[
   \left(\pi_n^a\right)^2
  \right]^2 
  -\frac{a^2}{16F^4}\sum_{n,\mu}
  \left(\pi_n^a\right)^2\left(\pi_{n+\mu}^b\right)^2
  \left[
    \left(\pi_n^c\right)^2+\left(\pi_{n+\mu}^c\right)^2
  \right]
  \nonumber \\
 &&+\frac{a^2}{16F^4}\left(m^2a^2+8\right)\sum_n
  \left[
   \left(\pi_n^a\right)^2
  \right]^3
  +\cdots, \\
 S_{Jacob}^{lat}&=&\sum_n
  \left[
   -\half\frac{(\pi^a_n)^2}{F^2}
   -\frac{1}{4}\frac{\left[(\pi^a_n)^2\right]^2}{F^4}
   +\cdots
  \right],
  \label{jacobisqrt}
\end{eqnarray}
where we retain only the terms which contribute to the two- and
four-point Green functions up to including $\mathcal{O}(p^4/F^4)$. Note
that, because of the discretization, it is difficult to count the power
of momenta buried, say, in $1-\cos(ap)$.  Instead we count the power of
$1/F$. There are no terms with positive power of $F$.

\begin{figure}[tbp]
\begin{tabular}{cc}
\begin{minipage}[t]{0.45\linewidth}
\begin{center}
\includegraphics[width=0.7\linewidth,clip]{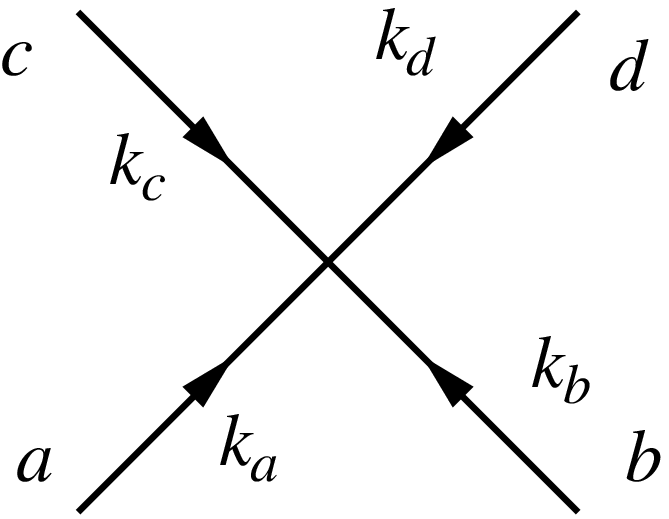}%
 \caption{\label{4-point_latt}Four-point vertex from
 $S_2^{lat}$. The indices $a,\cdots, d$ stands for the isospin of the
 pion field, and $k_a, \cdots, k_d$ are corresponding incoming momenta.}
\end{center}
\end{minipage}\quad\quad\quad
\begin{minipage}[t]{0.45\linewidth}
\begin{center}
 \includegraphics[width=0.7\linewidth,clip]{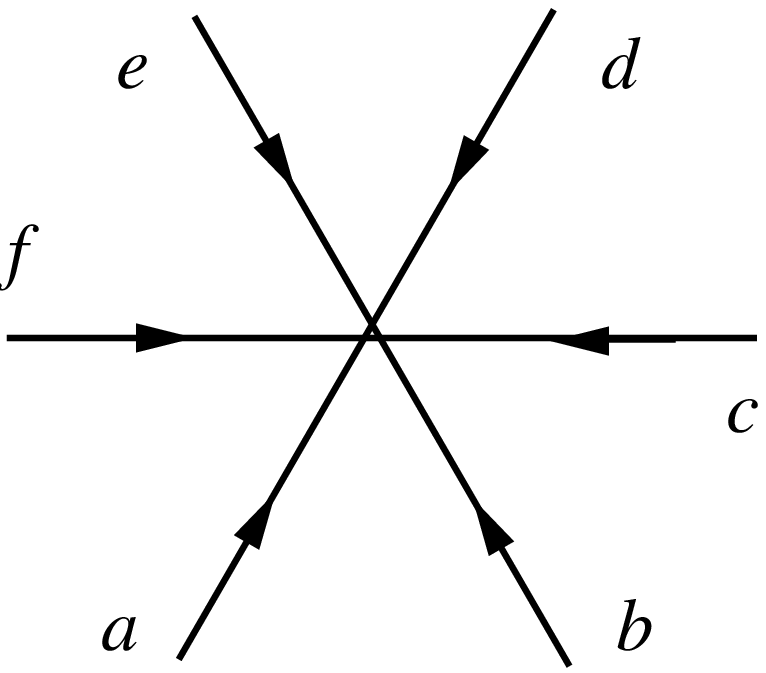}%
 \caption{\label{6-point_latt} Six-point vertex from $S_2^{lat}$. The
 indices $a,\cdots, f$ stands for the isospin of the pion field. The
 momentum labels are omitted.}
\end{center}
\end{minipage}
\end{tabular}
\end{figure}

The Feynman rules are obtained in the usual way, treating all the
contributions from $S_{Jacob}^{lat}$ as interactions. (They are of
higher order in $1/F$.) The propagator is the usual one,
\begin{equation}
 \langle \pi^a_n\pi^b_m\rangle_0 =
  \delta^{ab}\int_{\Box}\frac{d^4k}{(2\pi)^4}
  \frac{e^{ik(n-m)a}}
  {m^2 + \left[k\right]_a^2},
  \label{propagator}
\end{equation}
where $\int_\Box d^4k $ stands for the integration over the hypercube,
\begin{equation}
 \left\{
  k\ |\ k_\mu \in \left[ -\pi/a, \pi/a \right], \mu=1, \cdots, 4
 \right\},
\end{equation}
 and we have introduced a useful notation,
\begin{equation}
 \left[k\right]_a^2\equiv \frac{2}{a^2}\sum_\mu
  \left(1-\cos\left(k_\mu a\right)\right),
\end{equation}
which goes to $k^2$ in the continuum limit $a\rightarrow
0$. $S_2^{latt}$ leads to the following four-point and six-point
vertices:
\begin{equation}
 -\frac{1}{F^2}\sum_\mu 
  \bigg\{
  \delta^{ab}\delta^{cd}
  \left(
   [k_a+k_b]_a^2 +m^2
  \right)
  +
  \delta^{ac}\delta^{bd}
  \left(
   [k_a+k_c]_a^2 +m^2
  \right)
  +
  \delta^{ad}\delta^{bc}
  \left(
   [k_a+k_d]_a^2 +m^2
  \right)
  \bigg\},
\end{equation}
 and 
\begin{equation}
 -\frac{1}{F^4}
  \Bigg\{
  \delta^{ab}
  \left(
   [k_a+k_b]_a^2+ m^2
  \right)
  \left[
   \delta^{cd}\delta^{ef}
   \!\!+\!\delta^{ce}\delta^{df}
   \!\!+\!\delta^{cf}\delta^{de}
  \right]
  +14 \ \mbox{similar terms}
  \Bigg\},
\end{equation}
respectively. See FIG.~\ref{4-point_latt} and FIG.~\ref{6-point_latt}.

\subsection{Self-energy}

Shushpanov and Smilga~\cite{Shushpanov:1998ms} calculated the
self-energy contribution from the four-point vertex with a massless
propagator. We do the same calculation with a finite mass (See
FIG.~\ref{selfenergy_latt}),
\begin{equation}
 -\Sigma^{ab}(p) =-\delta^{ab}\Sigma(p)
  =-\frac{\delta^{ab}}{2F^2}\int_\Box \frac{d^4k}{(2\pi)^4}
  \frac{\left[k+p\right]_a^2+\left[k-p\right]_a^2+5m^2}
  {m^2+\left[k\right]_a^2}.
\end{equation}
Note that this leading order contribution is of order $1/F^2$.
Following their calculations, we find
\begin{equation}
 \Sigma(p)=
  \left[
   \frac{1}{2F^2a^2}\left(1+\frac{m^2a^2}{8}\right)\left[p\right]_a^2
   +\frac{3m^2}{4F^2a^2}
  \right]\mathcal{I}_0
  - \frac{1}{8F^2a^2}\left[p\right]_a^2
  +\frac{1}{F^2a^4},
  \label{selfenergy}
\end{equation}
where we have introduced $\mathcal{I}_0$,
\begin{equation}
 \mathcal{I}_n\equiv \int_0^\infty ds s^n
  e^{-\frac{s}{2}(m^2a^2+8)}\left[I_0(s)\right]^4,
  \label{commonint}
\end{equation}
and $I_0(s)$ is the modified Bessel function,
\begin{equation}
 I_0(s)=\int_{-\pi}^{\pi}\frac{dk}{2\pi}e^{s\cos k}.
\label{modifiedB}
\end{equation}

The last term of Eq.~(\ref{selfenergy}) is quartically divergent, and it
is cancelled by the $\mathcal{O}\left(1/F^2\right)$ contribution from
$S_{Jacob}^{lat}$, giving no ANTs. This cancellation mechanism is
well-known~\cite{Charap:1971bn, Honerkamp:1971va, Gerstein:1971fm}.

Note that modified Bessel function behaves for $s \gg 1$ as
\begin{equation}
 I_0(s) = \frac{e^s}{\sqrt{2\pi s}}
  \left(1+\mathcal{O}\left(\frac{1}{s}\right)\right),
\end{equation}
and for $0<s \ll 1$ as
\begin{equation}
 I_0(s) = 1+\mathcal{O}(s^2), 
\end{equation}
so that the integral $\mathcal{I}_n$ is finite as far as $m$ is kept
finite.  Although it is finite, but is not analytic at $m^2a^2=0$. One
cannot expand the result in terms of $m^2a^2$. This kind of singularity
at $m^2a^2=0$ persists in the calculations of the four-point functions
which we discuss in the next subsection. We therefore keep the mass terms
in the exponents (which come from the propagators) intact.

It is instructive to compare the cutoff integral (for $m \ll \Lambda$)
\begin{equation}
 2\pi^2\int_0^\Lambda \frac{k^3dk}{(2\pi)^4}\frac{1}{k^2+m^2}
  \sim c_2 \Lambda^2 + c_0 m^2 \ln \left(\frac{m^2}{\Lambda^2}\right),
\end{equation}
where $c_2$ and $c_0$ are numerical constants, with the corresponding
lattice version,
\begin{equation}
 \int_\Box \frac{d^4k}{(2\pi)^4}\frac{1}{[k^2]_a+m^2}
  =\frac{1}{2a^2}\mathcal{I}_0\, .
\end{equation}
By identifying $\Lambda\sim 1/a$, we see
\begin{equation}
 \mathcal{I}_0 \sim \tilde{c}_2 + \tilde{c}_0 (m^2a^2) \ln (m^2a^2)
\end{equation}
for $ma \ll 1$, where $\tilde{c}_2$ and $\tilde{c}_0$ are other numerical
constants. The second term causes the nonanalyticity of $\mathcal{I}_0$.
Similarly for $\mathcal{I}_1$, we have
\begin{equation}
 \mathcal{I}_1 \sim \tilde{d}_2 + \tilde{d}_0 \ln (m^2a^2)\, ,
\end{equation}
with some numerical constants $\tilde{d}_2$ and $\tilde{d}_0$.

\subsection{Four-point function}

\begin{figure}[tbp]
\begin{tabular}{cc}
\begin{minipage}[t]{0.45\linewidth}
\begin{center}
\includegraphics[width=0.7\linewidth,clip]{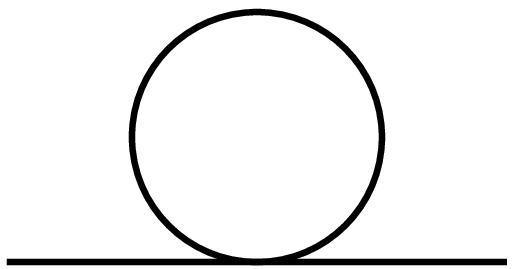}%
 \caption{\label{selfenergy_latt}Self-energy contribution from the
 four-point vertex from $S_2^{lat}$.}
\end{center}
\end{minipage}\quad\quad\quad
\begin{minipage}[t]{0.45\linewidth}
\begin{center}
 \includegraphics[width=0.7\linewidth,clip]{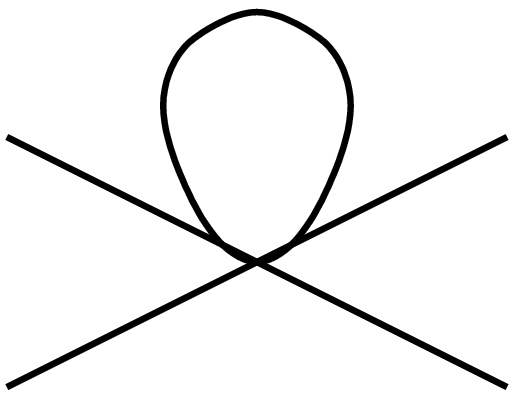}%
 \caption{\label{4-p_from6_latt}Contribution from the six-point vertex
 from $S_2^{lat}$ to the four-point function.}
\end{center}
\end{minipage}
\end{tabular}
\end{figure}

\begin{figure}[t]
 \includegraphics[width=0.8\linewidth,clip]{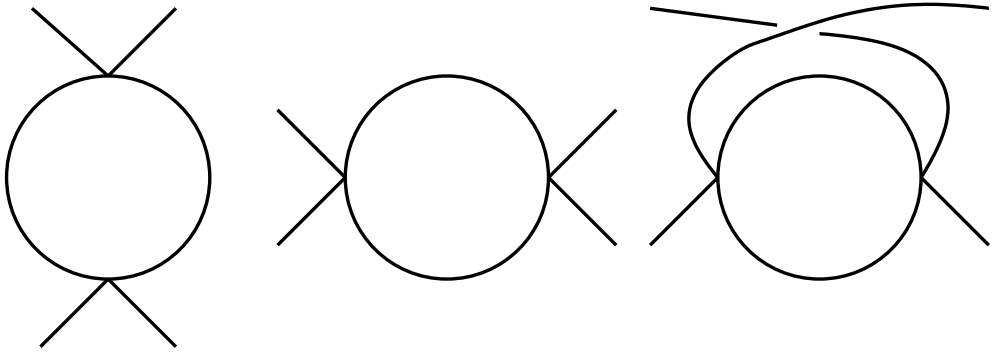}%
 \caption{\label{fish_latt}Three s-, t-, u-channel contributions from
  the four-point vertex from $S_2^{latt}$ to the four-point function.}
 \end{figure}

There are two kinds of contributions to the four-point function besides
the ones from $S_{Jacob}^{lat}$: the ones involving a six-point vertex
(FIG.~\ref{4-p_from6_latt}) and the ones involving two four-point
vertices (FIG.~\ref{fish_latt}).

In general, the four-point function in the continuum has the following
structure,
\begin{equation}
 \delta_{ab}\delta_{cd}A(p_a,p_b,p_c,p_d)
  +\delta_{ac}\delta_{bd}A(p_a,p_c,p_b,p_d)
  +\delta_{ad}\delta_{bc}A(p_a,p_b,p_d,p_c).
  \label{amp}
\end{equation}
It has the same structure on the lattice. Since the amplitude is
symmetric under the crossing, it is sufficient to calculate only the
contributions $A_L(p_a,p_b,p_c,p_d)$ on the lattice that correspond to
the first term $A(p_a,p_b,p_c,p_d)$ of Eq.~(\ref{amp}).


From FIG.~\ref{4-p_from6_latt}, we have the contribution 
\begin{eqnarray}
 A_L(p_a,p_b,p_c,p_d)^{\mbox{\scriptsize FIG.}~\ref{4-p_from6_latt}}
  \!\!
  &=&-\frac{1}{2F^4}\int_\Box \frac{d^4k}{(2\pi)^4}\frac{1}{m^2+[k]_a^2}
  \nonumber \\
 &&{}\times
  \Bigg\{
  10[p_a+p_b]_a^2
  +\!\!\!\!
  \sum_{i=a,b,c,d}\!\!\!\left([k+p_i]_a^2+[k-p_i]_a^2\right)
  \!+\!21m^2
  \Bigg\},
\end{eqnarray}
and from FIG.~\ref{fish_latt}, 
\begin{align}
 &A_L(p_a,p_b,p_c,p_d)^{\mbox{\scriptsize FIG.}~\ref{fish_latt}}
  \nonumber \\
  &=
  \frac{1}{2F^4} 
  \int_\Box\frac{d^4k}{(2\pi)^4}
  \frac{1}{m^2+\left[k\right]_a^2}
  \frac{1}{m^2+\left[p_a+p_b-k\right]_a^2}
  \nonumber \\
 &{} \times
  \bigg(
  3\left(\left[p_a+p_b\right]_a^2+m^2\right)^2
  +2\left(\left[p_a\!+\!p_b\right]_a^2\!+\!m^2\right)
  \left(\left[k\!+\!p_d\right]^2\!+\!\left[k\!-\!p_a\right]^2\!+\!2m^2\right)
  \bigg)
  \nonumber \\
 &{}+\frac{1}{2F^4} 
  \int_\Box\frac{d^4k}{(2\pi)^4}
  \frac{1}{m^2+\left[k\right]_a^2}
  \frac{1}{m^2+\left[p_a+p_c-k\right]_a^2}
  \nonumber \\
 &{} \times 
  2\left(\left[p_a-k\right]_a^2+m^2\right)
  \left(\left[k+p_b\right]_a^2+m^2\right)
  \nonumber \\
 & {}+\frac{1}{2F^4} 
  \int_\Box\frac{d^4k}{(2\pi)^4}
  \frac{1}{m^2+\left[k\right]_a^2}
  \frac{1}{m^2+\left[p_a+p_d-k\right]_a^2}
  \nonumber \\
 &{} \times 
  2\left(\left[p_a-k\right]_a^2+m^2\right)
  \left(\left[k+p_b\right]_a^2+m^2\right).
\end{align}
Note that these are of $1/F^4$.

If we set all the external momenta and the mass $m$ to be zero, we have 
\begin{eqnarray}
 \left. 
  A_L(p_a,p_b,p_c,p_d)^{\mbox{\scriptsize FIG.}~\ref{4-p_from6_latt}}
 \right|_{p_i=m=0}
  &=&-\frac{4}{F^4a^4}\, , \\
 \left. 
 A_L(p_a,p_b,p_c,p_d)^{\mbox{\scriptsize FIG.}~\ref{fish_latt}}
 \right|_{p_i=m=0}
  &=&\frac{2}{F^4a^4}\, .
\end{eqnarray}
The sum of them exactly cancels the $\mathcal{O}\left(1/F^4\right)$
contribution from the Jacobian, $2/F^4a^4$. Thus the amplitude satisfies
the soft-pion theorem.  There is no momentum (or mass) independent
ANT. Note that all the Jacobian contributions are used up to cancel the
momentum (and mass) independent contributions to this order. The
vertices from the Jacobian are now shown not to produce ANTs.

A straightforward but tedious calculation leads to the following result
for the one-loop contributions,
\begin{eqnarray}
 A_L(p_a,p_b,p_c,p_d)\!\!
  &=& \!
  -\frac{3\mathcal{I}_0}{4F^4a^2}(2s+3m^2)
  +\frac{s}{8F^4 a^2}\left(1-\half(8+m^2a^2)\mathcal{I}_0\right)
  \nonumber \\
 &&{}
  +\frac{\mathcal{I}_1}{24F^4}
  \left[
   9(s+m^2)^2-3(s+m^2)(2s-\Delta)+2 Z(p_a, p_b, p_c, p_d)
  \right]
  \nonumber \\
 &&{}
  +\frac{\mathcal{I}_0}{288F^4}
  \bigg[
   9(s+m^2)(2s-\Delta)-8 Z(p_a, p_b, p_c, p_d) 
   \nonumber \\
 &&{}
  \qquad \qquad \qquad \qquad \qquad \qquad \ \ 
   +48\sum_\mu(p_a)_\mu (p_b)_\mu (p_c)_\mu (p_d)_\mu,
  \bigg],
  \label{A_L_expanded}
\end{eqnarray}
where $s=(p_a+p_b)^2$, $t=(p_a+p_c)^2$, and $u=(p_a+p_d)^2$,
expanded in powers of the external momenta up to including
$\mathcal{O}(p^4/F^4)$. Here we have introduced the notation,
\begin{eqnarray}
 \Delta &\equiv& s + t + u\, , 
  \nonumber \\
 Z(p_a, p_b, p_c, p_d)
  &\equiv&
  \half
  \left[
   s(t+u)+2(t^2+u^2)-2(t+u)\Delta 
   +2(\Delta_{ac}\Delta_{bd}+\Delta_{ad}\Delta_{bc})
   -\Delta_{ab}\Delta_{cd}
  \right]\, ,
  \nonumber \\
 \Delta_{ij}&\equiv&
  p_i^2+p_j^2\, .
\end{eqnarray}
Some useful formulae to calculate Eq.~(\ref{A_L_expanded}) are given in
Appendix~\ref{sec:formulae}.

The terms proportional to $1/a^2$ correspond to quadratically divergent
ones. The chiral logarithms are contained in $\mathcal{I}_n$.  The last
term in Eq.~(\ref{A_L_expanded}) is not rotational invariant. It is not
a surprise, because the lattice regularization breaks rotational
invariance. 

In order to see if the result is manifestly chiral invariant, we need to
relate the expression to local operators. 

The terms in the first line of Eq.~(\ref{A_L_expanded}) are proportional
to $1/a^2$ (i.e., quadratically divergent) and quadratic in external
momenta. It is important to notice that they depend only on $s$ except
for the mass $m$. Note that there is only one chiral invariant operator
of $\mathcal{O}(p^2)$; Eq.~(\ref{lowest}) in the continuum. It produces
terms of exactly the same form as those in the first line, thus may
cancel the divergence. That is, the terms in the first line do not
contain ANTs.

A vigilant reader may notice that we have already considered the same
counterterm to cancel the divergence in the self-energy contribution,
thus its coefficient has been fixed. Here comes an important feature of
the perturbation theory; in terms of $U$, there is only one parameter,
i.e., the coupling constant $F$. On the other hand, when we introduce
the pion field, we have another parameter, the wave function
renormalization constant. Introducing the renormalized coupling constant
$F_R$ and the renormalized field $\pi^a_{Rn}$, we have 
\begin{equation}
 \frac{\pi^a_n}{F}=
  \left(\frac{1+\delta_\pi}{1+\delta_F}\right)
  \frac{\pi^a_{Rn}}{F_R}.
\end{equation}
By tuning only the parameter $\delta_\pi$, one can cancel the divergence
in the self-energy contribution. The parameter $\delta_F$ is now
determined to cancel the divergence in the first line of
Eq.~(\ref{A_L_expanded}).

Note that we consider the continuum action in order to see if ANTs
emerge. In momentum space, the difference between the continuum and the
lattice regularized ones is of higher order in momenta, and is not
rotational invariant. In order to cancel the divergence coming from the
difference, we need more counterterms which are of higher order in
momenta. Since they are not rotational invariant, the existence of such
counterterms do not interfere with the following argument for the
existence of ANTs, which, as we will see shortly, are rotational
invariant.

The terms in the second and third lines of Eq.~(\ref{A_L_expanded}) are
quartic in momenta (and the mass). The terms in the second line contain
logarithmic divergence due to $\mathcal{I}_1$, while those in the third
line are finite. There are only three chiral invariant operators of
$\mathcal{O}(p^4)$ available in the continuum;
\begin{eqnarray}
 \mathcal{O}_{1}&=&
\Tr\left(\partial_{\mu}U^{\dagger}\partial_{\mu}U\right)
\Tr\left(\partial_{\nu}U^{\dagger}\partial_{\nu}U\right),
\label{invop1}
\\
\mathcal{O}_{2}&=&
\mbox{Tr}\left(\partial_{\mu}U^{\dagger}\partial_{\nu}U\right)
\mbox{Tr}\left(\partial_{\mu}U^{\dagger}\partial_{\nu}U\right),
\label{invop2}
\\
\mathcal{O}_{3}&=&
\mbox{Tr}\left(\partial^2_{\mu}U^{\dagger}\partial^2_{\nu}U\right).
\label{invop3}
\end{eqnarray}
(Note that for $SU(2)$ there are some nontrivial relations which reduce
the number of independent operators. For example, $\Tr\left[(\der_\mu
U^\dagger \der_\mu U)^2 \right]$ is proportional to $\mathcal{O}_1$. )
If the terms in the second and third lines of Eq.~(\ref{A_L_expanded})
are of the same form as those produced by some linear combinations of
these operators, then these divergences may be cancelled by manifestly
chiral invariant operators. Let $C_i(p_a,p_b,p_c,p_d)/F^4\ (i=1,2,3)$
denote the contributions of these operators to the amplitude,
$A_L(p_a,p_b,p_c,p_d)$, to $\mathcal{O}(p^4/F^4)$. They are given by
\begin{eqnarray}
C_{1}(p_a,p_b,p_c,p_d)&=&
\left(s-\Delta_{ab}\right)\left(s-\Delta_{cd}\right),
\\
C_{2}(p_a,p_b,p_c,p_d)&=& 
\left(t-\Delta_{ac}\right)\left(t-\Delta_{bd}\right)
+
\left(u-\Delta_{bc}\right)\left(u-\Delta_{ad}\right),
\\
C_{3}(p_a,p_b,p_c,p_d)&=&
s^2,
\end{eqnarray}
respectively. In the massless limit, the terms in the square bracket in
the second line of Eq.~(\ref{A_L_expanded}) may be written as
\begin{equation}
 -C_{1}(p_a,p_b,p_c,p_d)
+2C_{2}(p_a,p_b,p_c,p_d)
+3C_{3}(p_a,p_b,p_c,p_d)
+3s\Delta,
\label{secondline}
\end{equation}
and those in the third line as
\begin{equation}
 4C_{1}(p_a,p_b,p_c,p_d)
-8C_{2}(p_a,p_b,p_c,p_d)
+18C_{3}(p_a,p_b,p_c,p_d)
-9s\Delta.
\label{thirdline}
\end{equation}
It is important to note that the last terms of Eqs.~(\ref{secondline})
and (\ref{thirdline}) cannot be expressed as a contribution of chiral
invariant operators. We have thus established the existence of ANTs. 

We remark that the terms which correspond to the logarithmic divergence,
Eq.~(\ref{secondline}), are different from those in the continuum.
Compare Eq.~(\ref{secondline}) with Eq.~(3.3) in
Ref.~\cite{Appelquist:1980ae}.

It is interesting to note that the ANTs are rotational invariant. We
also note that these are proportional to $\Delta$, i.e., the ANTs vanish
if the (massless) on-shell conditions are imposed for all the external
momenta.

The terms in the fourth line of Eq.~(\ref{A_L_expanded}) are
finite. They are manifestly chiral invariant, though they are not
rotational invariant. Actually, they can be obtained from the chiral
invariant operator of the form,
\begin{equation}
 \sum_\mu
  \Tr
  \left(
   \der_\mu U^\dagger \der_\mu U \der_\mu U^\dagger \der_\mu U
  \right).
\end{equation}
Even though it is uneasy to have such a rotational non-invariant term,
it has nothing to do with ANTs.

\section{Conclusion}
\label{sec:conclusion}

In this paper, we have established the existence of ANTs in lattice
chiral perturbation theory. Since the definition of the partition
function regularized on a lattice is manifestly chiral invariant (up to
the mass which regularizes the infrared singularities), and the
calculations are consistent with chiral symmetry, the symmetry is not
broken at all. Nevertheless the one-loop diagrams generate ANTs.  ANTs
are compatible with chiral symmetry.  The existence has been known in
the literature. Our contribution is the first demonstration of it in the
explicit lattice calculation.

On a lattice the Jacobian is well regularized, and we have shown that it
is not responsible for the appearance of ANTs. The role played by the
Jacobian is just to cancel the momentum independent, chirally
non-invariant contributions of the first kind mentioned in Introduction.

The result of the present paper has also given support for that the
appearance of ANTs is independent of regularization scheme. 

We find that the ANTs vanish when all the external momenta are on-shell,
consistent with the results obtained with dimensional regularization. It
means that the ANTs do not contribute to the S-matrix for the two-pion
scattering at least at the one-loop level.

Finally, we discuss a few points concerning ANTs, which are still
unclear to us.

Our original motivation for this study is related to setting up the
Wilsonian renormalization group calculation for the nonlinear sigma
model. The appearance of ANTs would cause a problem to the standard
program of the approach, even though they are compatible with chiral
symmetry. It would be desired to have a better statement of symmetry
than just the manifest invariance of the Wilsonian effective action. In
other words, we should seek for the combination of the Wilsonian program
and the Ward-Takahashi identities.

It is not clear to us if the ANTs in general (i.e., in higher order,
and/or in $n(>4)$-point functions) do not contribute to the S-matrix.
Ferrari \textit{et al.}~\cite{Ferrari:2005va} discussed general forms of
ANTs in the effective action, which is the generating function of the
one-particle irreducible Green functions. In order to see how these
terms contribute to the S-matrix, one needs to examine the effects of
one-particle reducible diagrams.

\appendix
\section{Some integration formulae}
\label{sec:formulae}

In this Appendix, we give some useful integration formulae for the
evaluation of $A_{L}(p_a,p_b,p_c,p_d)$ up to and including
$\mathcal{O}(p^4/F^4)$ discussed in Sec.~\ref{sec:lattice}.  

The basic technique that we make use of is Schwinger parameterization of
the propagator,
\begin{equation}
\frac{1}{m^2+[k]^2_a}
=
\int_0^{\infty}
ds\;e^{-s
\left(
 m^2+[k]^2_a
\right)}.
\label{schwing}
\end{equation}
To illustrate the method, let us consider the simple example,
\begin{equation}
 \int_{\Box} \frac{d^4 k}{(2\pi)^4}
\frac{1}{m^2+[k]_a^2}
\frac{1}{m^2+[k+p]_a^2}.
\end{equation}
By using Eq.(\ref{schwing}), it can be written as
\begin{equation}
 \int_0^\infty \!du \int_0^\infty \!dv 
  \int_{\Box} \frac{d^4 k}{(2\pi)^4}\;
  e^{-u\left(m^2+[k]_a^2\right)}
  e^{-v\left(m^2+[k+p]_a^2\right)}.
\end{equation}
Here we insert the identity,
\begin{equation}
1=\int_{0}^{\infty}ds\;
\delta\left(s-u-v\right),
\end{equation}
and making a change of variables, $v=s\alpha$, $k\rightarrow k/a$, and
$s\rightarrow sa^2/2$, we have
\begin{equation}
 \frac{1}{4}
  \int_0^1 \!d\alpha \int_{0}^{\infty}\!\! s\; ds 
  \int_{-\pi}^{\pi} \frac{d^4 k}{(2\pi)^4}\;
  e^{-s\left(M-\sum_\mu \cos k_\mu\right)}
  e^{s\alpha \left(\sum_\mu \cos(k_\mu+p_\mu a) -\sum_\mu \cos k_\mu\right)},
\end{equation}
where $M\equiv \left( \frac{8+m^2 a^2}{2} \right)$ is introduced.

In this way, all the necessary integrals may be written as the form,
\begin{equation}
 \int_0^\infty ds \; e^{-sM} \langle \langle X(p,k) \rangle\rangle,
\end{equation}
where we have introduced a useful notation $\langle \langle X(p,k)
\rangle\rangle$,
\begin{eqnarray} 
\langle \langle X(p,k)\rangle\rangle 
&\equiv&
\int_{-\pi}^{\pi}\frac{d^4 k}{(2\pi)^4} e^{s\sum_{\mu}\cos k_{\mu}}
X(p,k), \label{dblangle2}
\end{eqnarray}
with $X(p,k)$ being a function of the external momentum $p$ and the
dimensionless (i.e., rescaled) loop momentum $k$.

The diagrams we are interested in contain either a single propagator or
two propagators. For those involving a single propagator, the following
two integrals are relevant;
\begin{eqnarray}
\int_{\Box} \frac{d^4 k}{(2\pi)^4}
\frac{1}{m^2+[k]_a^2}
&=&
\frac{1}{2a^2}
\int_0^{\infty} ds
\; 
e^{-s M}
\langle\langle
1
\rangle\rangle,
\label{10}
\\
\int_{\Box} \frac{d^4 k}{(2\pi)^4}
\frac{1}{m^2+[k]_a^2}
\biggl(
m^2+[k+p]_a^2
\biggr)
&=&
\frac{1}{a^4}\int_0^{\infty}ds\;
e^{-s M}
\langle\langle
M-\sum_{\mu}\cos(k_{\mu}+p_{\mu}a) 
\rangle\rangle.
\label{11}
\end{eqnarray}

There are three types of integral that are relevant for one-loop
diagrams involving two propagators;
\begin{eqnarray}
&&\int_{\Box} \frac{d^4 k}{(2\pi)^4}
\frac{1}{m^2+[k]_a^2}
\frac{1}{m^2+[k+p]_a^2}
=
\frac{1}{4}
\int_0^{1}\hspace{-0.2cm}d\alpha 
\int_0^{\infty}\hspace{-0.3cm} s\; ds
 \; e^{-s M}
\langle \langle e^{-s \alpha N(p,k)}\rangle\rangle,
\label{20}
\\
&&\int_{\Box} \frac{d^4 k}{(2\pi)^4}
\frac{1}{m^2+[k]_a^2}
\frac{1}{m^2+[k+p]_a^2}
\biggl(
m^2+[k+q]_a^2
\biggr)
\nonumber \\
&&\quad=
\frac{1}{2a^2}
\int_0^{1}\hspace{-0.2cm}d\alpha
\int_0^{\infty}\hspace{-0.3cm} s \; ds 
\;e^{-s M}
\left\langle\!\! \left\langle 
\;e^{-s \alpha N(p,k)}
\biggl(
M-\sum_{\mu}\cos\left(k_{\mu}+q_{\mu} a\right)
\biggr)
\right\rangle\!\!\right\rangle,
\label{21}
\\
&&\int_{\Box} \frac{d^4 k}{(2\pi)^4}
\frac{1}{m^2+[k]_a^2}
\frac{1}{m^2+[k+p]_a^2}
\biggl(
m^2+[k+q]_a^2
\biggr)
\biggl(
m^2+[k+l]_a^2
\biggr)
\nonumber\\
&&\quad =
\frac{1}{a^4}\!
\int_0^{1}\hspace{-0.2cm}d\alpha\!\!
\int_0^{\infty}\hspace{-0.3cm} s\; ds\!
\;e^{-s M}
\left\langle\!\!\left\langle
e^{-s \alpha N(p,k)}
\biggl(
\!
M\!-\!\sum_{\mu}\cos\left(k_{\mu}\!+\!q_{\mu}a\right)
\!
\biggr)
\biggl(
\!
M\!-\!\sum_{\nu}\cos\left(k_{\nu}\!+\!l_{\nu} a\right)
\!
\biggr)
\right\rangle\!\! \right\rangle,
\nonumber\\
\label{22}
\end{eqnarray}
where $N(p,k)$ is defined as
\begin{eqnarray}
N(p,k)
&\equiv&
\sum_{\mu}
\Biggl[
\biggl(
1-\cos(p_{\mu}a)
\biggr)
\cos k_{\mu}+
\sin(p_{\mu}a)\sin k_{\mu}
\Biggr].
\end{eqnarray}
We can calculate $\langle \langle \quad \rangle\rangle$'s , by expanding
$e^{-s\alpha N(p,k)}$ in powers of external momenta and using the
following formulae,
\begin{align}
\langle\langle 1\rangle\rangle 
&=I^4_0
\\
\langle\langle \cos k_{\mu}\rangle\rangle 
&=I'_0 I^3_0
\\
\langle\langle \cos k_{\mu}\cos k_{\nu}\rangle\rangle
&=
\delta_{\mu\nu}
\biggl(I_0^4-\frac{1}{4s}
\left(I_0^4\right)'
-I_0^2 \left(I'_0\right)^2
\biggr)
+I_0^2 \left(I'_0\right)^2
\\
\langle\langle \sin k_{\mu}\sin k_{\nu}\rangle\rangle
&=
\frac{1}{4s}\left(I_0^4\right)'\delta_{\mu\nu}
\\
\langle\langle \cos k_{\mu}\sin k_{\nu} \sin k_{\lambda}\rangle\rangle
&=
\frac{1}{s}\delta_{\nu\lambda}
\Biggl[
\delta_{\mu\nu}
\biggl(
I_0^4-\frac{1}{2s}\left(I^4_0\right)'
-I_0^2 \left(I'_0\right)^2
\biggr)
+
I_0^2 \left(I'_0\right)^2
\Biggr]
\\
\langle \langle 
\sin k_{\mu}\sin k_{\nu}\sin k_{\lambda}\sin k_{\gamma}
\rangle\rangle
&=
\frac{1}{s^2}\delta_{\mu\nu}\delta_{\lambda\gamma}
\biggl[
\delta_{\mu\lambda}
\biggl(
I_0^4-\frac{1}{2s}\left(I_0^4\right)'-I_0^2\left(I'_0\right)^2
\biggr)
+I_0^2\left(I'_0\right)^2
\biggr]
\nonumber\\
&+
\frac{1}{s^2}\delta_{\mu\lambda}\delta_{\nu\gamma}
\biggl[
\delta_{\mu\nu}
\biggl(
I_0^4-\frac{1}{2s}\left(I_0^4\right)'-I_0^2\left(I'_0\right)^2
\biggr)
+I_0^2\left(I'_0\right)^2
\biggr]
\nonumber\\
&+
\frac{1}{s^2}\delta_{\mu\gamma}\delta_{\nu\lambda}
\biggl[
\delta_{\mu\lambda}
\biggl(
I_0^4-\frac{1}{2s}\left(I_0^4\right)'-I_0^2\left(I'_0\right)^2
\biggr)
+I_0^2\left(I'_0\right)^2
\biggr],
\end{align}
where $I_0(s)$ is the modified Bessel function given in
Eq.~(\ref{modifiedB}).  The prime stands for a derivative with respect
to $s$. Note that a bracket $\langle\langle\cdots\rangle\rangle$
containing an odd number of ($\sin k_\mu$)'s vanishes because of parity.

It is important to notice that all integrands in Eqs.~(\ref{20}),
(\ref{21}), and (\ref{22}) contain the exponential suppression factor
$e^{-s M}$ with $M > 4$.  It justifies the expansion of $e^{-s\alpha
N(p,k)}$ in powers of $(ap_{\mu})$ within the integrals even though
$N(p,k)$ is multiplied by $s$, since it effectively cuts off the domain
of integration where $s$ is large.

Now Eqs.~(\ref{20}), (\ref{21}), and (\ref{22}) can be expressed in
terms of $\mathcal{I}_n$ defined in Eq.~(\ref{commonint}). In doing so,
we extensively use the identity
\begin{equation}
I^{''}_0(s)
=
I_0(s)-\frac{1}{s}I^{'}_0(s),
\end{equation}
which is nothing but the modified Bessel differential equation satisfied
by $I_0(s)$.

Finally we obtain the integrals involving a single propagator,
\begin{eqnarray}
&&\int_{\Box} \frac{d^4 k}{(2\pi)^4}
\frac{1}{m^2+[k]_a^2}
=
\frac{1}{2a^2}\mathcal{I}_0\;, 
\\
&&\int_{\Box} \frac{d^4 k}{(2\pi)^4}
\frac{1}{m^2+[k]_a^2}
\biggl(
m^2+[k+p]_a^2
\biggr)
=
\frac{1}{a^4}
\biggl[
M\mathcal{I}_0+\frac{1}{4}
\left(
1-M\mathcal{I}_0
\right)
\sum_{\mu}\cos(p_{\mu}a)
\biggr], 
\end{eqnarray}
and those involving two propagators,
\begin{eqnarray}
&&\int_{\Box} \frac{d^4 k}{(2\pi)^4}
\frac{1}{m^2+[k]_a^2}
\frac{1}{m^2+[k+p]_a^2}
=
\frac{1}{4}
\Biggl[
\mathcal{I}_{1}
+O((ap)^2)
\Biggr]\;, 
\\
&&\int_{\Box} \frac{d^4 k}{(2\pi)^4}
\frac{1}{m^2+[k]_a^2}
\frac{1}{m^2+[k+p]_a^2}
\biggl(
m^2+[k+q]_a^2
\biggr)
\nonumber\\
&&\quad =
\frac{1}{2a^2}
\biggl[
\mathcal{I}_0
+
\frac{1}{8}
\sum_{\mu}a^2 \left(p_{\mu}q_{\mu}-q^2_{\mu}\right)
\left(
\mathcal{I}_0-M \mathcal{I}_1
\right)
+O((ap)^4)
\biggr],
\nonumber\\
&&\int_{\Box} \frac{d^4 k}{(2\pi)^4}
\frac{1}{m^2+[k]_a^2}
\frac{1}{m^2+[k+p]_a^2}
\biggl(
m^2+[k+q]_a^2
\biggr)
\biggl(
m^2+[k+l]_a^2
\biggr)
\nonumber\\
&&\quad =
\frac{1}{a^4}
\Biggl[
1
-\frac{a^2}{8}
[p-(q+l)]_a^2
\left(
1-M\mathcal{I}_0
\right)
\nonumber\\
&&\quad +
\frac{1}{4}\mathcal{I}_1
\sum_{\mu}
a^4
\biggl\{
p^2_{\mu}q_{\mu}l_{\mu}
-p_{\mu}\left(q_{\mu}l^2_{\mu}+l_{\mu}q^2_{\mu}\right)
+q^2_{\mu}l^2_{\mu}
\biggr\}
\nonumber\\
&& \quad
+\frac{1}{144}
\left(
M\mathcal{I}_0+(4-M^2)\mathcal{I}_1
\right)
\sum_{\mu,\nu}a^4
\Biggl\{
3p^2_{\mu}q_{\mu}l_{\mu}
+p^2_{\mu}(q_{\nu}l_{\nu})
-4(p_{\mu}q_{\mu})(p_{\nu}l_{\nu})
\nonumber\\
&&\quad 
-3p_{\mu}\left(q_{\mu}l^2_{\mu}+l_{\mu}q^2_{\mu}\right)
+3(p_{\mu}q_{\mu})l^2_{\nu}
+3(p_{\mu}l_{\mu})q^2_{\nu}
+3\left(q^2_{\mu}l^2_{\mu}-q^2_{\mu}l^2_{\nu}\right)
\Biggr\}
+O((ap)^6)
\Biggr],
\end{eqnarray}
where only the necessary terms to calculate $A_L(p_a,p_b,p_c,p_d)$ to order
$\mathcal{O}(p^4/F^4)$ are retained.

\begin{acknowledgments}
 The authors are grateful to Atsushi Ninomiya for the discussions. The
 discussions with Hiroshi Yoneyama are also acknowledged.
\end{acknowledgments}

\bibliography{NLS,NEFT}

\end{document}